\documentclass[preprint,preprintnumbers,amsmath,amssymb,superscriptaddress]{revtex4}
\usepackage{graphicx}
\usepackage{bm}
\usepackage{amsmath}
\usepackage{amsfonts}
\usepackage{amssymb}%
\usepackage{float}
\usepackage{stmaryrd}

\newcommand{\be}{
\begin{equation}
}
\newcommand{\ee}{
\end{equation}
}
\newcommand{\beq}{
\begin{eqnarray}
}
\newcommand{\eeq}{
\end{eqnarray}
}
\begin{document}
\title{Cover times of  random searches\\
 }

\author{Marie Chupeau}
\affiliation{Laboratoire de Physique Th\'eorique de la Mati\`ere Condens\'ee, UMR 7600 CNRS /UPMC, 4 Place Jussieu, 75255
Paris Cedex, France}

\author{Olivier B\'enichou\footnote{benichou@lptmc.jussieu.fr}}
\affiliation{Laboratoire de Physique Th\'eorique de la Mati\`ere Condens\'ee, UMR 7600 CNRS /UPMC, 4 Place Jussieu, 75255
Paris Cedex, France}

\author{Rapha\"el Voituriez\footnote{voiturie@lptmc.jussieu.fr}}
\affiliation{Laboratoire de Physique Th\'eorique de la Mati\`ere Condens\'ee, UMR 7600 CNRS /UPMC, 4 Place Jussieu, 75255
Paris Cedex, France}
\affiliation{Laboratoire Jean Perrin, UMR 8237 CNRS /UPMC, 4 Place Jussieu, 75255
Paris Cedex}

\date{\today}

\begin{abstract}

How long does it take a random searcher to visit all sites of a given domain? This time, known as the cover time \cite{Aldous:1983ai},   is a key observable to quantify the efficiency of exhaustive searches, which require a complete exploration of an area and not only the discovery of a single target; examples range from  immune system cells chasing pathogens \cite{Heuze:2013uq}  to animals harvesting resources \cite{Viswanathan:2008,Benichou:2011fk}, robotized exploration by e.g. automated cleaners or deminers, or algorithmics \cite{Vergassola:2007}.   
Despite its broad relevance, the cover time has remained  elusive and so far explicit results  have been scarce and mostly limited to regular random walks \cite{Brummelhuis:1991ys,Dembo:2004zr,ding,Belius:2013xy}.
Here we determine the full distribution of the cover time for a broad range of random search processes, which includes the prominent examples of  L\'evy strategies \cite{shles_klaft,Viswanathan:1996,Viswanathan:1999a,R.Metzler:2000,Lomholt:2008}, intermittent strategies \cite{Benichou:2005qd,Benichou:2011fk,Oshanin:2007a},  persistent random walks \cite{Tejedor:2012ly} and random walks on complex networks \cite{Condamin:2007zl},  and reveal its universal features.   We show that for all these examples the mean cover time can be minimized, and that the corresponding  optimal strategies  also minimize the mean search time for a single target,  unambiguously pointing towards their robustness.

\end{abstract}

\maketitle
\begin{figure}[H]
\includegraphics[width=13cm]{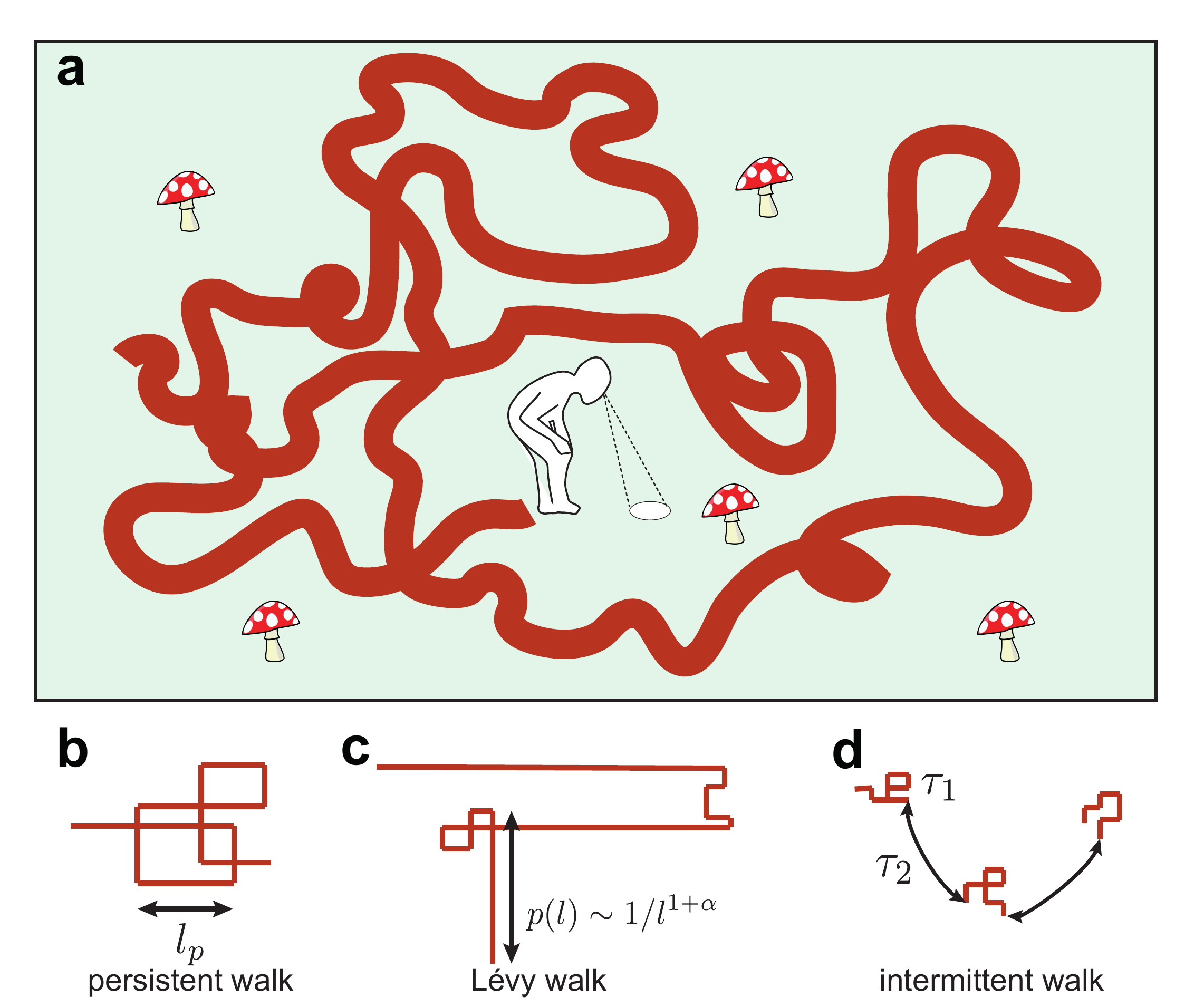}
\caption{
}
\label{fig0}
\end{figure}

Random search processes have proved over the recent years to be involved in a broad range of contexts at various scales,  from the search of specific sequences on DNA by proteins to animal foraging \cite{Viswanathan:2008,Benichou:2011fk}. So far,  the main tools to quantify the efficiency of such search processes could be expressed in terms of the  time needed for the searcher to reach a single target, the so--called first-passage time \cite{Redner:2001a,Condamin:2007zl,BenichouO.:2010,Benichou:2014fk,Bray:2013}. However, as soon as several targets need to be found, which is a recurrent situation in chemistry, ecology, or robotics,   the relevant observable is rather  the time needed to  reach a fraction of the domain sites (see Fig. \ref{fig0}). The extreme case  of such exhaustive searches where all sites of a domain need to be visited defines the so-called cover time, which is of particular interest since it gives  the time needed to find all targets of a domain with certainty; its    determination is a long standing problem of  random walk theory \cite{Aldous:1983ai,Aldous:1989uq} (see \cite{Weiss:1982rz, Burov:2007} for related observables).

Nevertheless, analytical results on cover times are scarce. Important steps were achieved in \cite{Yokoi:1990kx}, where the mean cover time of an interval was analytically calculated for one-dimensional symmetric nearest-neighbor random walks, both for periodic and reflecting boundary conditions. In dimensions greater or equal to three, Aldous \cite{Aldous:1989uq} has determined the leading behavior of the mean cover time in the limit of large domain size, which was reproduced by numerical simulations in \cite{Nemirovsky:1990vn}. In the physics literature, these results have been extended  to the two-dimensional case in \cite{Brummelhuis:1991ys}, which has been since then refined in the mathematics literature \cite{Dembo:2004zr,ding,Belius:2013xy}. Notably, all these results were so far essentially limited to the case of regular random walks, i.e. symmetric nearest-neighbor random walks in Euclidean  geometries.

 However, recently, several classes of more complex random search strategies, including L\'evy strategies \cite{shles_klaft,Viswanathan:1996,Viswanathan:1999a,R.Metzler:2000,Lomholt:2008}, intermittent strategies \cite{Benichou:2005qd,Benichou:2011fk,Oshanin:2007a} and persistent random walks \cite{Tejedor:2012ly}, have emerged and been shown theoretically to be efficient (see Fig. \ref{fig0}). In this context, existing theoretical studies have up to now  focused mainly on the first-passage time to a single target, and the cover time of these random search processes, required to quantify the efficiency of exhaustive searches, has been left aside. The analytical determination of  the entire distribution of cover time type observables  for general classes of random walks, including the above,  is at the core of this paper.

We consider a random walker moving on a  network of $N$ sites.  The random walk is assumed to be Markovian and non compact (i.e.  transient: in infinite space, the  probability for the walker to ever reach a  given site is strictly smaller than one \cite{Hughes:1995}).  This  covers a large class of processes relevant to search problems. We denote by $\tau(M,N)$ the partial cover time, defined as the time needed to visit {\it any} $M$ distinct sites of the network. The alternative problem of determining the time needed to visit $M$ {\it given} sites (chosen at random), called the random cover time  \cite{Nemirovsky:1990vn}, will be addressed below. Note that taking $M=N$ (for both the partial and the random cover time) yields the full cover time, to which most of the literature has been devoted so far. In practice, we will focus on large values of $M,N$. The starting point is to introduce $\theta(N-k,N)\equiv \tau(k+1,N)-\tau(k,N)$, defined as the time needed for the number of distinct sites visited by the random walker to increase from $k$ to $k+1$. Alternatively,  $\theta(N-k,N)$ is the time needed to visit a new site among the $N-k$ unvisited sites, once $k$ sites have been visited.  The following exact expression then holds:
\begin{equation}
\label{step_distribution}
\tau(M,N)=\sum_{k=N-M+1}^{N-1} \theta(k,N).
\end{equation}
The exact determination of the statistical properties of this random variable can seem out of reach since a priori $\theta(k,N)$ depends on  the entire random trajectory until time \mbox{$\tau(N-k,N)$}. 
The key hypothesis is then to assume that  the random variables $\theta(k,N)$ are in fact  independent asymptotically in the large $N$ limit. This hypothesis will be verified numerically for  all tested non compact random walks, and will furthermore allow to retrieve  exact results known so far for  regular random walks. In this regime of $N$ large the distribution of $\theta(k,N)$ can be obtained explicitly and  enables the determination of the full distribution of the partial cover time, which, as we show in Supplementary Information (SI), finally takes the universal form
\begin{equation}\label{dist}
P(x)=\frac{1}{p!}\exp(-(p+1)x-e^{-x})
\end{equation}
valid in the limit $N,M\to\infty$ with $p\equiv N-M$ fixed (implying in particular $p/N\to 0$). Here the rescaled variable  \mbox{$x\equiv\tau/\langle T\rangle-\ln N$} involves the mean  $\langle T\rangle$ of the global  first-passage time, defined as the mean first-passage time to a given target site averaged over all starting sites. This constitutes the central result of this paper. Its derivation  relies on the fact that the distribution of the global first-passage time $T$ to  a given target site  is asymptotically an exponential of mean $\langle T\rangle$ for non compact random walks \cite{Meyer:2011,Benichou:2011}. In addition, we assumed that $\langle T\rangle$ is independent of the target site, which is exact for domains with periodic boundary conditions, and in practice well satisfied for any domain shape in the case of non compact random walks, as was checked numerically. 

Several comments are in order. (i) The result of equation  (\ref{dist}) unveils the universal dependence of the distribution of the partial cover time on both the random walk process -- through only the global mean first-passage time to a single target -- , and the geometry of the domain -- through only its volume $N$ (using that $\langle T\rangle$ asymptotically depends on the geometry only through  $N$ \cite{Condamin:2007zl,Benichou:2014fk}).
(ii) The result of equation  (\ref{dist}) reveals a deep connexion with order statistics, already pointed out in the mathematical literature for regular random walks  \cite{Belius:2013xy}. Indeed, equation  (\ref{dist}) is  the limit distribution   of the $(p+1)^{th}$ largest among $N$ independent identically distributed random variables in the  large $N$ regime (Gumbel universality class). In fact, in the case $M=N$, the full cover time can easily be seen as the largest among the  first-passage times $t_i$  of the searcher to site $i$, where $ i\in \llbracket 1,N \rrbracket$ covers all sites of the domain.  equation  (\ref{dist}) indeed yields in this case ($p=0$) the classical Gumbel law. More generally, 
 the partial cover time $\tau(N-p,N)$ can be seen as the $(p+1)^{th}$ largest among the $\{t_i\}_{i\in \llbracket 1,N \rrbracket}$. What we find here is that in the case of non compact exploration, the $\{t_i\}_{i\in \llbracket 1,N \rrbracket}$ are asymptotically independent in the  large $N$ regime. (iii) The result of equation  (\ref{dist}) provides in particular an explicit determination of the mean partial cover time  $\langle \tau\rangle\sim \langle T\rangle (\ln(N)-\Psi^{(0)}(p+1))$ in the  large $N$ regime, where $\Psi^{(n)}$ denotes the polygamma function of order $n$, and which  is in agreement with exact results obtained for the full cover time in the particular case of Brownian walks on periodic lattices \cite{Aldous:1989uq,Belius:2013xy}. (iv) Beyond the mean, the variance of the partial cover time can be determined and is given asymptotically by $\sigma^2 \sim \Psi^{(1)}(p+1) \langle T\rangle^2$. This corresponds to a reduced variance $\sigma^2/\langle \tau\rangle^2\sim \Psi^{(1)}(p+1)/\ln^2 N$, which shows that the amplitude of the relative fluctuations slowly decays with the domain size. (v) The distribution of related observables can be readily deduced from this approach, and depends on the search process only through the global mean  first-passage time $\langle T\rangle$ to a given target in the domain for a single searcher.  For example the distribution of the random cover time, defined above as the time needed to visit $M$ {\it given} sites of a domain of $N$ sites  \cite{Nemirovsky:1990vn},  is obtained in the limit  $M,N\gg 1$ by taking $p=0$ in equation  (\ref{dist}) (Gumbel law), with the new rescaled variable $x\equiv \tau/\langle T\rangle-\ln M$ (see SI). In turn, in the important case where $n$ searchers explore the domain simultaneously,  the distribution of the partial cover time  is also given by equation  (\ref{dist}), with the new rescaled variable $x\equiv n\tau/\langle T\rangle-\ln N$.

\begin{figure}[H]
\includegraphics[width=18cm]{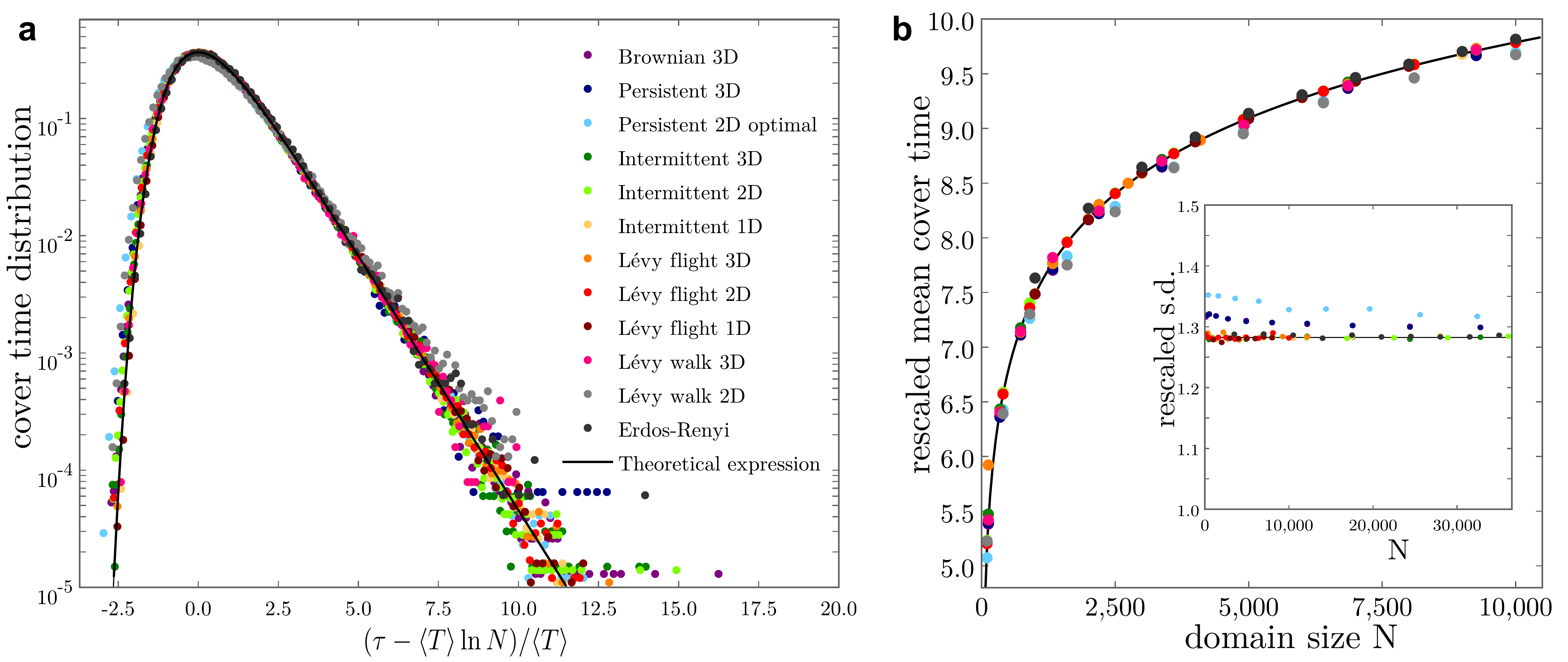}
\caption{\vspace{5cm}}
\label{fig1}
\end{figure}

\begin{figure}[H]
\includegraphics[width=18cm]{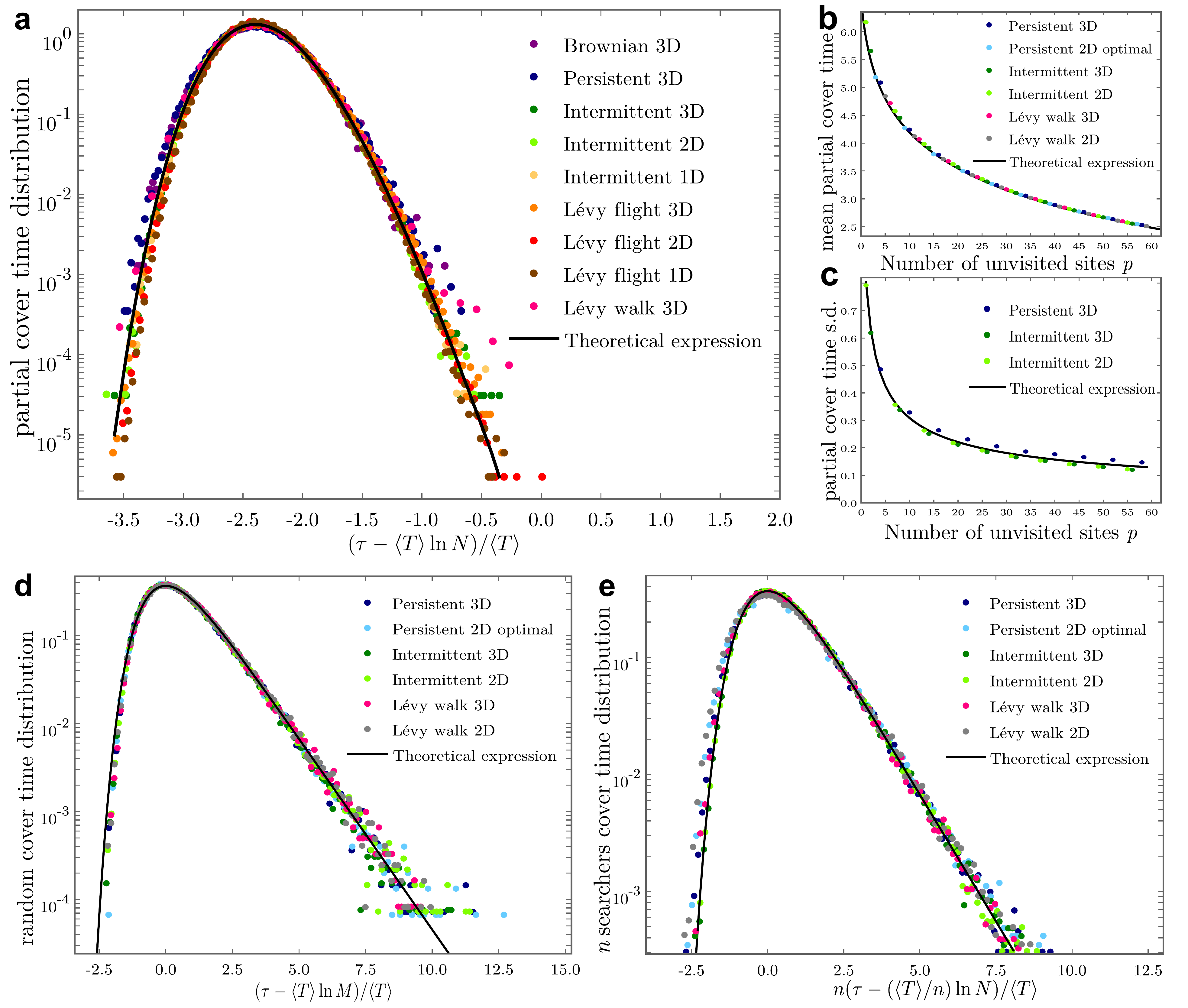}
\caption{\vspace{5cm}}
\label{fig2}
\end{figure}

We now confirm these analytical results by Monte Carlo simulations of most of the  models of  random search strategies invoked in the literature:  Brownian random walks, persistent random walks,  L\'evy flights and L\'evy walks,  intermittent random walks, and random walks on complex networks. 
Figures (\ref{fig1}) and (\ref{fig2})  reveal excellent quantitative agreement between the analytical predictions and the numerical simulations. The prediction of equation   (\ref{dist})  unambiguously captures the mean, variance, and entire distribution of partial and full cover times (Fig. (\ref{fig1}) and (\ref{fig2})), as well as random cover times and cover times for $n$ searchers (Fig. (\ref{fig2})), as shown by the data collapse of the numerical simulations. We emphasize that the very different nature of these examples demonstrates that the range of applicability of our approach, which mainly relies on the non compact property of the random trajectory of the searcher, is wide. Note that even for $2D$ persistent random walks, which  are  in fact marginally compact, both the mean and variance of  cover times are quantitatively predicted by our approach in the large domain size limit, provided that the persistence length minimizes the global mean first-passage time for a target $\langle T\rangle$,  as analyzed in  \cite{Tejedor:2012ly}.

\begin{figure}[H]
\includegraphics[width=18cm]{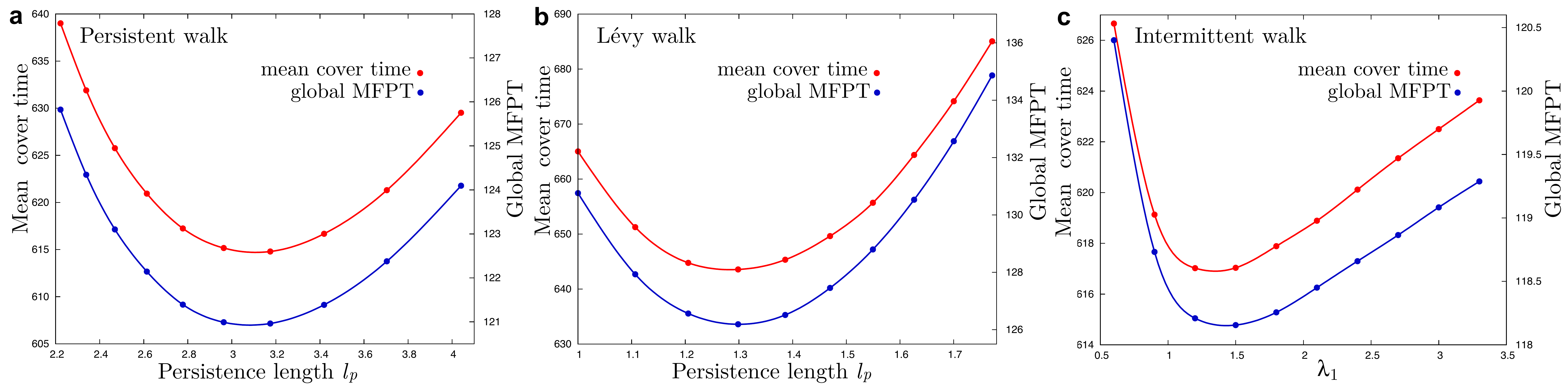}
\caption{}
\label{fig3}
\end{figure}

In practice, the question of minimizing the cover time in order to optimize the efficiency of the search process is crucial. Our analysis reveals that the mean cover time (either random or partial)  is minimized exactly when the  global  mean first-passage time  $\langle T\rangle$ for a {\it single target} is minimized (see Fig. (\ref{fig3})). In particular, in the case of persistent random walks and intermittent random walks, which have been shown to minimize $\langle T\rangle$ (either by tuning the persistence length of the persistent random walk or the duration of the scanning phase of the intermittent random walk), we find that mean  cover times can also be minimized.  For L\'evy walks, the celebrated optimization of the target encounter rate  for a  L\'evy index $\alpha\simeq 1$ (defined through the jump length distribution $p(l)\propto l^{-\alpha-1}$ for $l$ large) has been obtained and discussed only for a  distribution of infinitely many so-called revisitable targets (which reappear at the same position after being found) \cite{Viswanathan:1999a}. In contrast, the global  mean first-passage time to a single target in confinement, which by definition involves a non revisitable target, has been left aside. In fact, we find that $\langle T\rangle$ (see also \cite{Tejedor:2012ly}), and therefore the mean cover time, can be minimized by adjusting the persistence length for all $\alpha>1$ (see Fig. (\ref{fig3})). This  optimal strategy, which does not require revisitable targets,   is therefore very different from the above mentioned  optimal strategy obtained for  infinitely many  revisitable targets. All together, these results  shed new light on the role of persistent, intermittent and L\'evy strategies in the optimization of search processes, and clearly points towards their  robustness.

\section*{ Methods}
 The definition of random search processes analyzed in the text are as follows (see SI for details):\\
 {\it (i) Brownian random walks} constitute the most striking example of random search process, which is known to be non compact in dimension $D=3$, and marginally compact for $D=2$. Here we consider nearest neighbor random walks on a periodic lattice of $N$ sites. \\
 {\it (ii) L\'evy flights} have been shown to play an important role in random search problems \cite{R.Metzler:2000}. We consider here  discrete  L\'evy flights of index $\alpha$ on $1D$, $2D$ and $3D$ 
periodic lattices of size $N$, characterized by a probability distribution to perform a jump of size $l$ that obeys $p(l)\propto l^{-\alpha-1}$ for $l$ large.  \\
{\it (iii) Complex networks.} Beyond classical Euclidean spaces, many examples of random walks on complex networks, whose relevance to extremely various fields is now unanimously recognized,   are non compact. Here we consider the emblematic Erd\H{o}s-R\'enyi networks \cite{Albert:2001RMP}.\\
{\it (iv) Persistent random walks}  provide a minimal example of search process with memory, which is encoded in the persistence length, defined as the  mean number of successive steps performed in a given direction. They have been shown to lead to a minimization of the mean search time for a single target, and therefore play a prominent role in the optimization of search processes \cite{Tejedor:2012ly}
. We consider here $2D$ and $3D$ discrete persistent random walks on a periodic lattice of $N$ sites.\\
{\it (v) L\'evy walks}. As opposed to L\'evy flights, which can have arbitrary large velocities, L\'evy walks \cite{shles_klaft} have a constant speed, and can be seen as an extension of persistent random walks, for which the distribution of the number of successive steps is not exponential, but power-law distributed. They have been shown to be optimizable and have been extensively invoked in the context of animal behavior \cite{Viswanathan:1999a}. We consider here $2D$ and $3D$ discrete L\'evy walks on a periodic lattice of $N$ sites.\\
{\it  (vi) Intermittent strategies.} Last, we consider the case of  well hidden targets, for which it can be assumed that  moving and searching are incompatible. In this case,  the search strategy is intermittent \cite{Benichou:2011fk}. Technically, we focus here on a  continuous time two-state searcher moving on a $1D$, $2D$ or  $3D$ periodic lattice of $N$ sites \cite{Benichou:2007fv}. In the slow reactive state 1, the searcher performs a regular nearest neighbor random walk with jump rate $\rho$, and actually visits the corresponding sites. With rate $\lambda_1\equiv 1/\tau_1$, the searcher switches to a fast and non reactive state 2, which enables a uniform relocalization in the domain, but during which no sites are visited. The searcher then switches to state 1 with rate $\lambda_2\equiv 1/\tau_2$. 

\section*{ Acknowledgements}
 O.B. was supported by  ERC grant FPTOpt-277998.
 
\section*{ Author contributions}
All authors contributed equally to this work\\

\newpage
{\bf FIGURE LEGENDS}\\

{\bf FIGURE 1}\\
How long does it take to exhaustively explore a given domain? This quantity defines the cover time of the domain.\\
{\bf a}. An example of exhaustive search: the time needed to find all mushrooms and exhaust a given area, with no prior knowledge of their distribution in space, is the cover time of the domain. In this paper we also consider the time needed to visit any $M$ sites of the domain, defined as the partial cover time, as well as the time needed to visit $M$ given sites of the domain chosen at random, defined as the random cover time. 
Examples of optimizable search processes (see Methods): {\bf b} persistent random walks, {\bf c} L\'evy walks, and {\bf d} intermittent random walks.
For all these search processes, we show that the distribution of the cover time takes a universal form, and that the mean cover time can be minimized.
\\

{\bf FIGURE 2}\\
Universal distribution of the full cover time ($M=N$) for non compact search processes. \\
{\bf a}. Distribution of the rescaled cover time for various non compact search processes. All data collapse to a universal master curve defined by equation (\ref{dist}) with $p=0$ (plain line).   {\bf b}. Mean cover time (rescaled by the global mean first-passage time) as a function of the domain size $N$. The plain line gives the theoretical prediction $\ln N-\Psi^{(0)}(1)=\ln N+\gamma$, where $\gamma$ denotes the Euler constant. Inset: standard deviation of the cover time (rescaled by the global mean first-passage time) as a function of the domain size $N$. The plain line gives the theoretical prediction $\Psi^{(1)}(1)=\pi^2/6$.\\
For all panels, domain sizes and all parameters defining the search processes are listed  in SI.  \\

{\bf FIGURE 3}\\
Universal distribution of cover time type observables for non compact search processes. \\
{\bf a}. Distribution of the rescaled partial cover time for various non compact search processes. All data collapse to a universal master curve defined by equation (\ref{dist}),  here for $p\equiv N-M=10$ unvisited sites (plain line).  {\bf b}. Mean partial cover time (rescaled by the global mean first-passage time) as a function of  $p$ for $N$ fixed. The plain line gives the theoretical prediction $\ln(N)-\Psi^{(0)}(p+1)$.  {\bf c.} Standard deviation of the partial cover time (rescaled by the global mean first-passage time) as a function of $p$ for $N$ fixed. The plain line gives the theoretical prediction $\Psi^{(1)}(p+1)$. {\bf d}.  Distribution of the rescaled random cover time for various non compact search processes. All data collapse to a universal master curve defined by equation (\ref{dist}) with $p=0$, here for $M=20$ randomly chosen given  sites to visit (plain line). {\bf e}.  Distribution of the rescaled full cover time for various non compact search processes with $n$ independent searchers. All data collapse to a universal master curve defined by equation (\ref{dist}) with $p=0$, here for $n=10$ searchers (plain line). \\
For all panels, domain sizes (all such that $N\gg 1$) and all parameters defining the search processes are listed  in SI. \\

{\bf FIGURE 4}\\
The mean full cover time and the mean search time for a single target can be minimized by the same optimal strategy.
The mean full cover time and the global mean first-passage time to a single target are plotted as a function of the persistence length for persistent random walks ({\bf a}, here in $2D$, $N=100$) and L\'evy walks ({\bf b}, here in $2D$ with $\alpha=1.8$, $N=100 $), and as a function of the switching rate $\lambda_1$ (see methods and SI) for intermittent random walks ({\bf c}, here with $\rho=1$,  $\lambda_2=0.8$, $N=100$). \\

\pagebreak

\pagebreak

\pagebreak

\pagebreak

\pagebreak

\begin{Large}
\begin{center}
Supplementary Information
\end{center}
\end{Large}

\section{definitions and derivation of the cover time distribution}
We consider a random walker moving on a  network of $N$ sites.  The random walk is assumed to be Markovian and non compact. 
\subsection{Partial cover time}
We denote by $\tau(M,N)$ the partial cover time, defined as the time needed to visit any $M$ distinct sites of the network. Note that taking $M=N$ yields the full cover time. In practice, we will focus on the regime  $M,N\gg 1$.  We first  introduce $\theta(N-k,N)\equiv \tau(k+1,N)-\tau(k,N)$, defined as the time needed for the number of distinct sites visited by the random walker to increase from $k$ to $k+1$. Alternatively,  $\theta(N-k,N)$ is the time needed to visit a new site among the $N-k$ unvisited sites, once $k$ sites have been visited.  The following exact expression then holds:
\begin{equation}
\label{step_distributionSI}
\tau(M,N)=\sum_{k=N-M+1}^{N-1} \theta(k,N).
\end{equation}
Note that a priori $\theta(k,N)$ depends on  the entire random trajectory until time $\tau(N-k,N)$, so that the $\theta(k,N)$ are not independent random variables.

We now argue that the distribution of $\theta(k,N)$ can be written in the regime $N\gg1$ and $k\ll N$:
\begin{equation}\label{fptdistSI}
f_{k,N}(\theta(k,N)=t)\sim\frac{k}{\langle T\rangle}\exp (-kt/\langle T\rangle),
\end{equation}
where $\langle T\rangle$ denotes the global mean first-passage time, i.e. the mean first-passage time to a target site, averaged over all starting positions of the random walker. Note that $\langle T\rangle$ encompasses the dependence on $N$, and only weakly depends on the position of the target site for non compact random walks.  To derive Eq.(\ref{fptdistSI}), we first  notice that by definition $\theta(k,N)$ is the first-passage time of the searcher to any of the $k$ unvisited sites, once $N-k$ sites have been visited. We next make use of the general large volume asymptotics of the first-passage time distribution to a single target for non compact random walks derived in  \cite{BenichouO.:2010SI,Meyer:2011SI}, which was shown to be a single exponential of mean $\langle T\rangle$ after averaging over the starting position. Last, it is assumed   that for $k\ll N$, the $k$ remaining unvisited sites are not clustered (to avoid screening effects, which are only short ranged for non compact processes \cite{Condamin:2007euSI,Benichou:2014fkSI} ). These $k$ unvisited sites can then be considered as independent,  which finally yields Eq.(\ref{fptdistSI}). This assumption implies that the $\theta(k,N)$ are in fact independent in the regime $N\gg1$ and $k\ll N$. Using Eq.(\ref{step_distributionSI}), the cover time can then be expressed asymptotically as a sum of independent random variables.
We then introduce the Laplace transform
\begin{equation}\label{fptdistlaplaceSI}
{\hat f}_{k,N}(s)\equiv\int_0^\infty dt \, e^{-st} f_{k,N}(t)=\frac{1}{1+s\langle T\rangle/k},
\end{equation}
and conclude that the Laplace transform  of the distribution $P(\tau)$ of the cover time can be written
\begin{equation}\label{distlaplaceSI}
{\hat P}(s)\sim\prod_{k=N-M+1}^{N-1}\frac{1}{1+s\langle T\rangle/k}=\frac{(N-1)!}{(N-M)!}\frac{\Gamma(N-M+1+s\langle T\rangle)}{\Gamma(N+s\langle T\rangle)}
\end{equation}
which yields in the regime $N\gg 1$, where $p\equiv N-M$ is kept fixed:
\begin{equation}\label{distlaplace2SI}
{\hat P}(s)\sim\frac{\Gamma(p+1+s\langle T\rangle)}{p!N^{s\langle T\rangle}}.
\end{equation}
Introducing the rescaled variable  $x\equiv\tau/\langle T\rangle-\ln N$, one finds after Laplace inversion:
\begin{equation}\label{distSI}
P(x)\sim\frac{1}{p!}\exp(-(p+1)x-e^{-x}),
\end{equation}
which is valid in the limit $N,M\to\infty$ with $p= N-M$ fixed (implying in particular that $p/N$ is small).
\begin{figure}
\includegraphics[width=13cm]{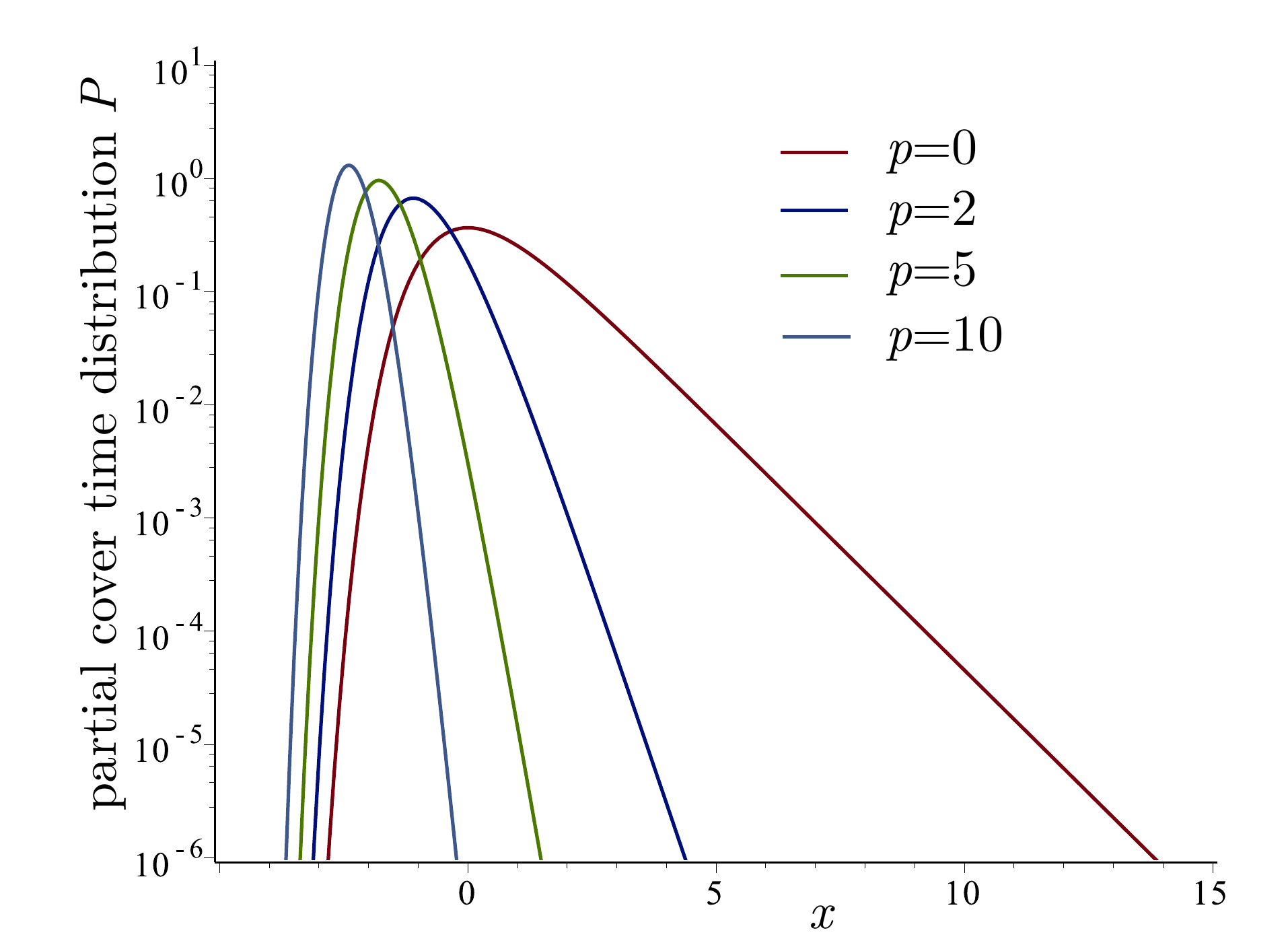}
\caption{Universal distribution of the partial cover time for different values of $p$, as given by Eq.\ref{distSI}.
}
\label{S1SI}
\end{figure}
 This function is plotted in Fig. S\ref{S1SI} for different values of $p$. In particular, for $p=0$ one finds the classical Gumbel law.  The first moments of the partial cover time can then be readily deduced:
\begin{equation}\label{mom1SI}
\langle\tau(M,N)\rangle\sim \langle T\rangle(\ln N -\Psi^{(0)}(p+1))
\end{equation}
\begin{equation}\label{mom2SI}
\langle\tau(M,N)^2\rangle-\langle\tau(M,N)\rangle^2\sim \langle T\rangle^2\Psi^{(1)}(p+1))
\end{equation}
where $\Psi^{(n)}(x)$ is the polygamma function of order $n$.
In particular, taking $p=0$  yields with $\Psi^{(0)}(1)=-\gamma$ and $\Psi^{(1)}(1)=\pi^2/6$  the exact results derived for regular random walks on the torus \cite{Brummelhuis:1991ysSI,Dembo:2004zrSI,dingSI,Belius:2013xySI}. Last we note that the case of $n$ independent searchers can be straightforwardly deduced by the substitution $ \langle T\rangle\to  \langle T\rangle/n$.

\subsection{Random cover time}
We now consider the case of the random cover time $\tau_r(M,N)$, defined as the time needed to visit $M$ given sites of interest (chosen at random) of the network of $N$ sites.
Following the above section, we write:
\begin{equation}
\label{step_distribution2SI}
\tau_r(M,N)=\sum_{k=1}^{M-1} \theta_r(k),
\end{equation}
where here $\theta_r(k)$ denotes the time needed to visit a new site of interest among the $k$ sites of interest that have not yet been visited (the dependence on $N,M$ is omitted for clarity).
Following the above section,  the distribution of $\theta_r(k)$ can be written in the regime $M\gg1$ (and therefore $N\gg 1$) and $k\ll M$:
\begin{equation}\label{fptdist2SI}
f_{k}^{(r)}(\theta_r(k)=t)\sim\frac{k}{\langle T\rangle}\exp (-kt/\langle T\rangle).
\end{equation}
Note that here, as above, $\langle T\rangle$ denotes the global mean first-passage time to a target site of the domain of $N$ sites, which depends on $N$ but not on $M$.  Following the above section, the Laplace transformed distribution of the random cover time can be written:
\begin{equation}\label{distlaplace3SI}
{\hat P}_r(s)\sim\frac{\Gamma(1+s\langle T\rangle)}{M^{s\langle T\rangle}}.
\end{equation}
Introducing the rescaled variable  $x\equiv\tau/\langle T\rangle-\ln M$, one finds after Laplace inversion:
\begin{equation}\label{dist3SI}
P_r(x)\sim \exp(-x-e^{-x}),
\end{equation}
which is valid in the limit $N,M\to\infty$. This is the classical Gumbel law.

\begin{figure}
\includegraphics[width=13cm]{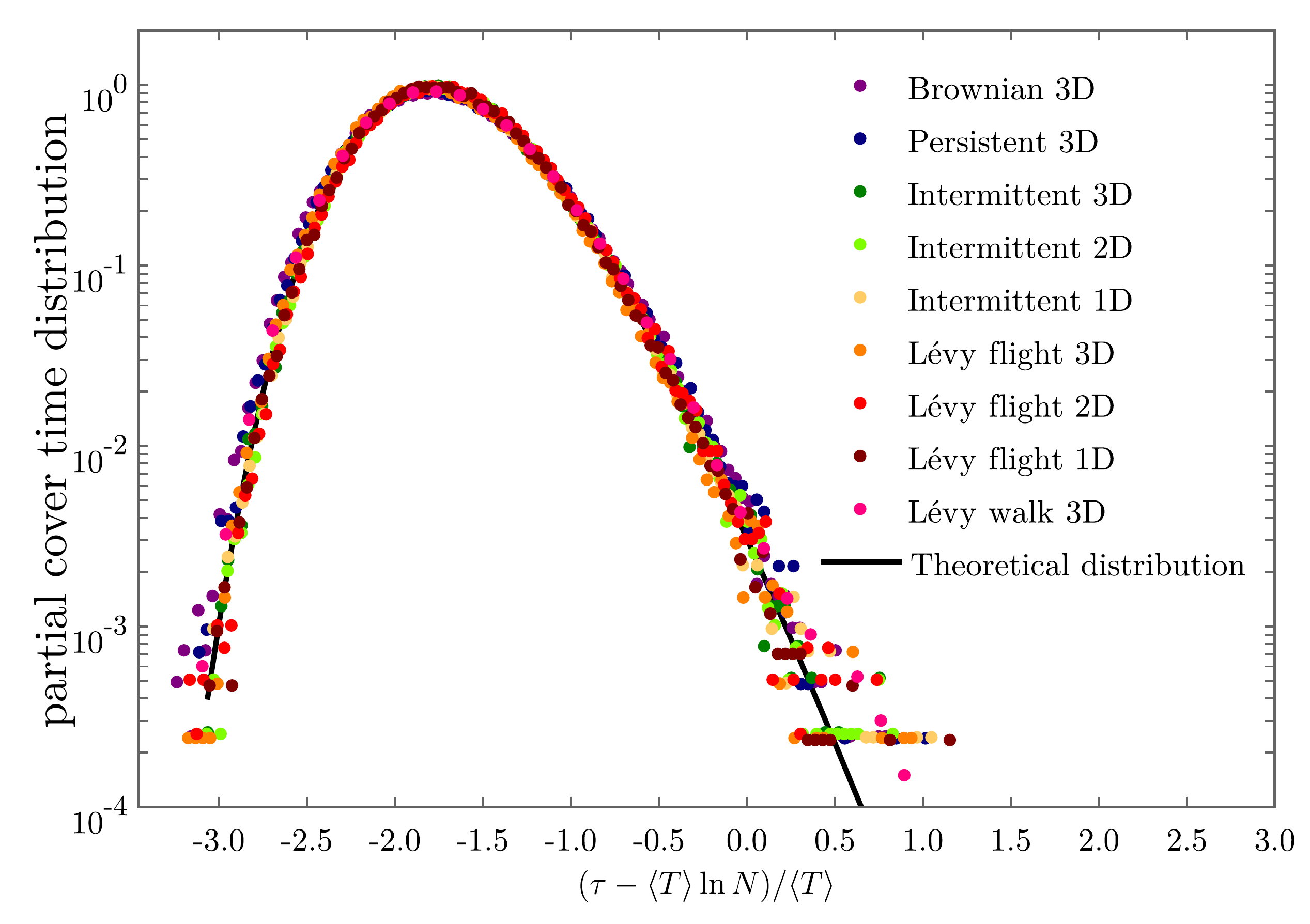}
\caption{Distribution of the rescaled partial cover time for various non compact search processes. All data collapse to a universal master curve defined by Eq.(\ref{distSI}),  here for $p=5$ unvisited sites (plain line). Other parameters as in Fig. 3 of the main text (see below).
}
\label{S2SI}
\end{figure}
\begin{figure}
\includegraphics[width=13cm]{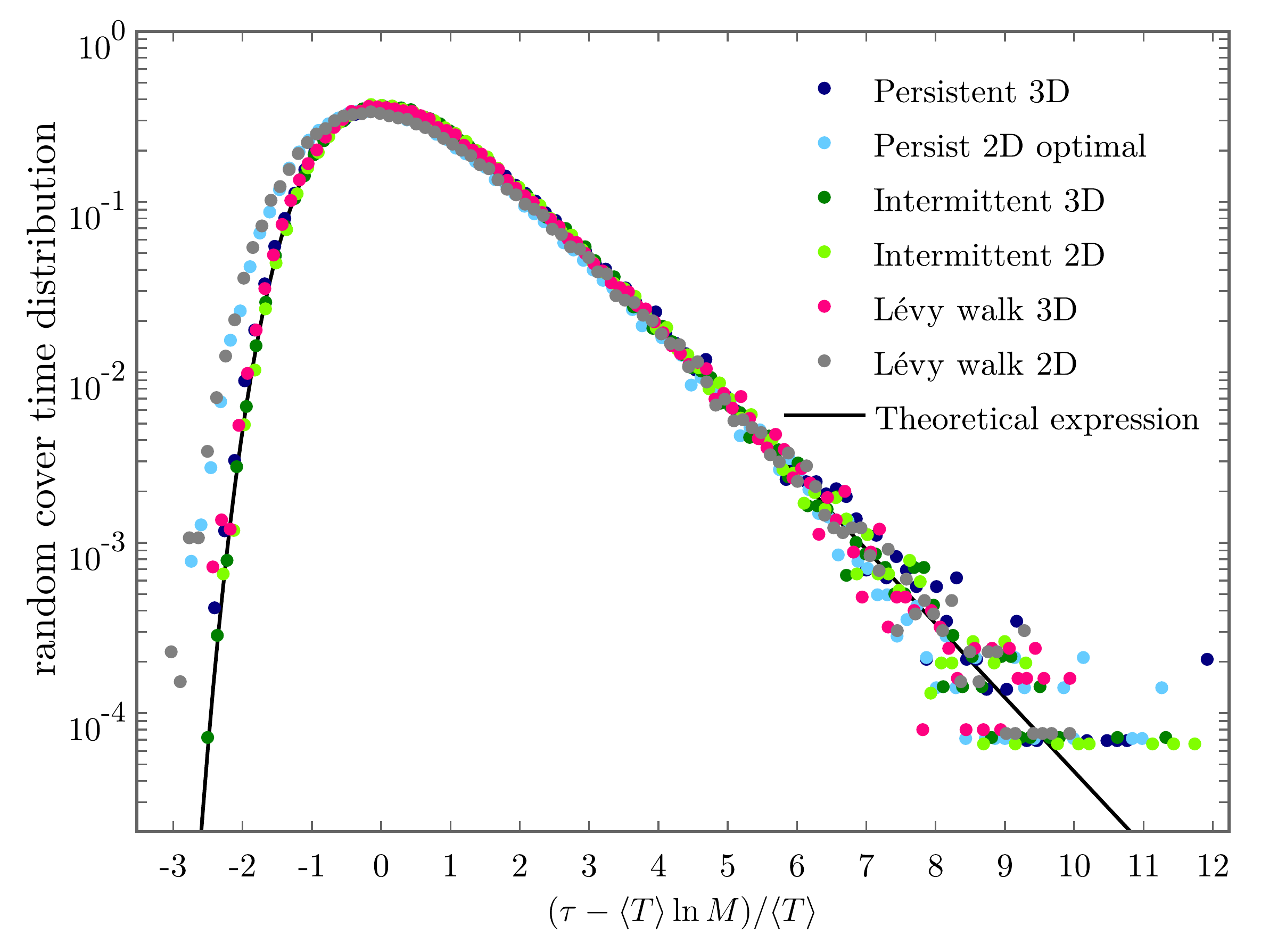}
\caption{Distribution of the rescaled random cover time for various non compact search processes. All data collapse to a universal master curve defined by Eq.(\ref{dist3SI}), here for $M=N-10$ randomly chosen given  sites to visit (plain line). Other parameters as in Fig. 3 of the main text (see below).
}
\label{S3SI}
\end{figure}

\section{Numerical simulations}
The analytical results of the previous section have been checked by numerical simulations of various non compact search processes as discussed below. The excellent agreement is discussed in the main text (Figs 2,3). Further examples are given in Figs. S\ref{S2SI} and S\ref{S3SI}.
\subsection{Definition of the search processes}
\begin{figure}
\includegraphics[width=13cm]{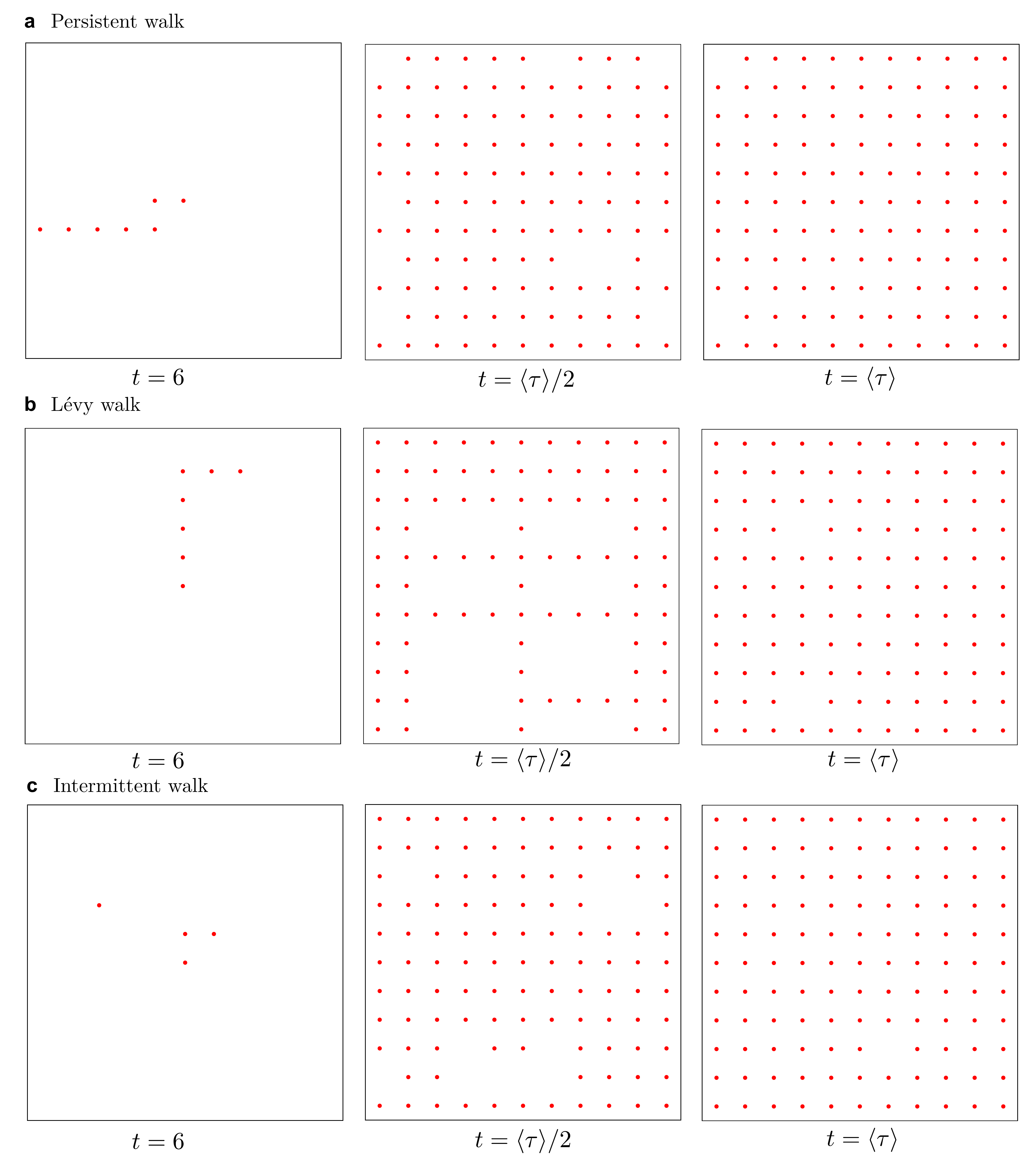}
\caption{Examples of covered territory for the three main classes of search processes at different time points. Persistent walk with $l_p=3.5$, L\'evy walk with $l_p=3.5$ and $\alpha=1.1$, intermittent walk with $\rho=1,\lambda_1=\lambda_2=0.1$
}
\label{S4SI}
\end{figure}

The random search processes defined in the main text are generated numerically as detailed below. Examples of trajectories are given in Fig. S\ref{S4SI}.

\textit{(i) Brownian random walks.} We consider discrete Brownian random walks on periodic lattices of $N$ sites in dimension $D$. At each time step, the random walker moves to one of its nearest neighbors with probability $1/(2D)$. 

\textit{(ii) L\'evy flights.} We consider discrete L\'evy flights  on periodic lattices of $N$ sites in dimension $D$. At each time step, the walker jumps in one of the $2D$ directions chosen randomly, the jump distance $l$ being drawn from a L\'evy distribution of exponent $\alpha$, defined by:
\begin{equation}\label{levySI}
p(l) =\frac{1}{\pi} \int_{-\infty}^{\infty}dk \; e^{ikl-l_0^\alpha |k|^\alpha}
\end{equation}
 for $l>0$, where $l_0$ is a scale parameter.  Since the distance $l$ is a real number, the arrival point is off lattice. In the numerical simulations,  the walker is therefore relocated  at the nearest site after each jump. Last, note that for each  jump, only the arrival site is considered as visited.

\textit{(iii) Erd\H{o}s-R\'enyi networks.} We consider a nearest neighbor random walk on the classical Erd\H{o}s-R\'enyi network of $N$ sites, defined as follows : for each pair of sites,  a link exists with a fixed probability $\nu$. The connectivity $c_i$ of each site therefore depends on the site $i$. At each time step, the walker jumps to one of the neighboring sites  with probability $=1/c_i$.

\textit{(iv) Persistent random walks.} We consider discrete persistent random walks  on periodic lattices of $N$ sites in dimension $D$. At each time step, the walker jumps to one of its nearest neighbors with probabilities  that depend on the previous step. The probability to go on in the same direction is $1/(2D)+\epsilon (2D-1)/(2D)$ whereas the probability to jump in one of the $2D-1$ other directions is $(1-\epsilon)/(2D)$. The persistence length, defined as the mean number of steps performed in the same direction, is then given by $l_p=\frac{2D}{(2D-1)(1-\epsilon)}$

\textit{(v) L\'evy walks.} We consider discrete L\'evy walks  on periodic lattices of $N$ sites in dimension $D$. The walker performs ballistic  excursions of random direction. The length  of each ballistic excursion (equal to its  number of steps) is drawn from  a L\'evy law of exponent $\alpha$ defined in Eq. (\ref{levySI}), and the direction chosen uniformly from the $2D$ directions of the lattice. Note  that as opposed  to L\'evy flights, the duration of each ballistic  excursion is by definition its number of steps, and that the walker  visits all the sites of a ballistic excursion, and not only the starting and final ones. When $\alpha>1$, the mean number of steps of each excursion is finite,  and one can identify the persistence length with the scale parameter  $l_p\equiv l_0$. 

\textit{(vi) Intermittent strategies.} We consider continuous time intermittent walks  on periodic lattices of $N$ sites in dimension $D$. At each time, the walker can either perform a step to one of its nearest neighbors at a rate $\rho$ or relocate anywhere on the lattice at a rate $\lambda_1$. The total rate of such events is therefore $\lambda_{\textrm{tot}}=\lambda_1+\rho$. Using the Gillespie algorithm \cite{Gillespie:1976SI}, the waiting time before the next event, which will be a diffusive step with probability $
\rho/\lambda_{\textrm{tot}}$ or a relocation step with probability $\lambda_1/\lambda_{\textrm{tot}}$, is drawn from an exponential law of rate $\lambda_{\textrm{tot}}$. In addition each relocation step takes a time drawn from an exponential distribution of rate $\lambda_2$.

\subsection{Parameters of the numerical simulations displayed in the figures of the main text }

\subsubsection{Figure 2}
{\bf a.} Distributions of the rescaled full cover time are plotted as a function of the rescaled cover time for 3D Brownian walks (here with a domain size $N=125000$), persistent random walks (here, in 3D with $N=6859$ and a persistence length $l_p=6$, and in 2D with $N=400$ and $l_p=4.8$), intermittent random walks (here, in 3D with $N=1331$, a step rate in \mbox{phase 1} $\rho=20$, a switch rate to \mbox{phase 2} $\lambda_1=20$, and a switch rate to \mbox{phase 1} $\lambda_2=5$; in 2D with $N=400$, $\rho=20$, $\lambda_1=10$, $\lambda_2=5$; in 1D with $N=100$, $\rho=0.1$, $\lambda_1=\lambda_2=1$), L\'evy flights (here in 3D with $N=1000$ and index $\alpha=1.5$; in 2D with $N=400$ and $\alpha=1.5$; in 1D with $N=100$ and $\alpha=0.5$, with a scale parameter $l_0=1$ for these three cases), L\'evy walks (here in 3D for $N=9261$ and in 2D for $N=100$, with $\alpha=1.8$ and a persistence length  $l_p=1.3$ for both cases) and for Erd\H{o}s-R\'enyi networks (with $N=10000$ and a link probability $\nu=0.3$).  {\bf b.} Mean cover time (rescaled by the global mean first-passage time) as a function of the domain size $N$, for 3D Brownian walks, persistent walks (here in 3D with $l_p=2.4$ and in 2D for the persistence length that minimizes the mean search time of one target among $N$ sites), intermittent walks (in 3D, 2D and 1D with $\rho=0.1$ and $\lambda_1=\lambda_2=1$), L\'evy flights (in 3D, 2D and 1D with $\alpha=0.1$ and $l_0=1$), L\'evy walks (in 3D for $\alpha=1.8$ and $l_p=1.3$, and in 2D for $\alpha=1.4$ and $l_p=1$) and Erd\H{o}s-R\'enyi networks (for $\nu=0.3$). Inset: standard deviation of the cover time (rescaled by the global mean first-passage time) as a function of the domain size $N$, for the same types of walks, except L\'evy walks for which the standard deviation is infinite.\\

\subsubsection{Figure 3}
{\bf a}. Distribution of the rescaled partial cover time with $p=N-M=10$ fixed for 3D Brownian walks (here with $N=10^6$), 3D persistent walks (with $N=9261$ and $l_p=2.4$), intermittent walks (in 3D with $N=1000$, $\rho=1$, $\lambda_1=5$ and $\lambda_2=20$; in 2D with $N=900$, and 1D with $N=1000$, and for the last two, with $\rho=0.1$ and $\lambda_1=\lambda_2=1$), L\'evy flights (for $\alpha=0.1$ and $l_0=1$, in 3D with $N=512$, in 2D with $N=400$ and in 1D with $N=500$) and L\'evy walks in 3D (with $N=29791$, $\alpha=1.8$ and $l_p=1.3$). {\bf b.} Mean partial cover time (rescaled by the global mean first-passage time) as a function of $p=N-M$ for $N=729$ fixed, for persistent walks (in 3D with $l_p=4.3$ and in 2D with $l_p=4.8$), intermittent walks (with $\rho=0.1$ and $\lambda_1=\lambda_2=1$) and L\'evy walks (with $\alpha=1.8$ and $l_p=1.3$). {\bf c.} Standard deviation of the partial cover time (rescaled by the global mean first-passage time) as a function of $p$ for $N$ fixed. The parameters are the same as in {\bf b.} except for 3D persistent walks (here $N=9261$ and $l_p=2.4$). {\bf d}.  Distribution of the rescaled random cover time for persistent walks for $M=20$ given sites chosen at random (in 3D with $N=29791$ and $l_p=2.4$, and in 2D with $N=441$ and $l_p=5.6$), intermittent walks (in 3D with $N=1331$, in 2D with $N=400$, both for $\rho=0.1$ and $\lambda_1=\lambda_2=1$) and L\'evy walks (in 3D with $N=9261$, in 2D with $N=10201$, both for $\alpha=1.8$ and $l_p=1.3$). {\bf e}. Distribution of the rescaled full   cover time with $n$ independent searchers, for persistent walks (in 3D with $N=1331$ and $l_p=2.4$, and in 2D with $N=441$ and $l_p=5.6$), intermittent walks (in 3D with $N=1331$ and in 2D with $N=961$, both for $\rho=0.1$ and $\lambda_1=\lambda_2=1$) and L\'evy walks (in 3D with $N=1331$ and in 2D with $N=961$, both with $\alpha=1.8$ and $l_p=1.3$).



\end{document}